\documentclass{article}
\usepackage{hiph-art}
\newcommand{\beq}{\begin{equation}}
\newcommand{\eeq}{\end{equation}}
\newcommand{\bea}{\begin{eqnarray}}
\newcommand{\eea}{\end{eqnarray}}
\newcommand{\bef}{\begin{figure}}
\newcommand{\eef}{\end{figure}}
\newcommand{\bce}{\begin{center}}
\newcommand{\ece}{\end{center}}
\newcommand{\eg}{{\it e.g.}}
\newcommand{\ie}{{\it i.e.}}
\newcommand{\etal}{{\it et al.}}
\def\lsim{\mathrel{\rlap{\lower4pt\hbox{\hskip1pt$\sim$}}
    \raise1pt\hbox{$<$}}}         
\def\gsim{\mathrel{\rlap{\lower4pt\hbox{\hskip1pt$\sim$}}
    \raise1pt\hbox{$>$}}}         

\def\rightmark{Electromagnetic Radiation and in-Medium Effects}
\def\leftmark{Ralf Rapp}
 
\title{Electromagnetic Radiation and in-Medium Effects} 
\authors{
{$Ralf Rapp$ %
\index{Rapp, R.} 
}\\[2.812mm]
{\normalsize
\hspace*{-8pt} Cyclotron Institute and Physics Department,
Texas A\&M University,\\ 
College Station, TX 77843-3366, USA\\[0.2ex] 
}}
 
\abstract{The theory of thermal photon and dilepton 
emission from a hot and dense hadronic gas, as well as from the
Quark-Gluon Plasma, is reviewed in the context of extracting
in-medium properties of the matter constituents.
In phenomenological applications to ultrarelativistic heavy-ion
collisions we focus on recent photon and dilepton spectra as 
measured by WA98 and CERES/NA45, respectively, at CERN-SPS energies.}

 
\makeindex
\begin{document}
 
\maketitle

\section{Introduction}
\label{intro}
Among the key objectives in studying the properties of a hot 
and/or dense strongly interacting medium is the identification 
of the relevant degrees of freedom that render the
most economic description of the pertinent phases of matter.
Of special interest are modes that can be connected
to (pseudo-) order parameters of the bulk matter. These
modes are expected to undergo substantial spectral modifications
in the vicinity of phase changes, and thus can be used as
indicators of the latter.

Spectacular experimental results have been obtained over the last
few years at the Relativistic Heavy-Ion Collider (RHIC) in $Au$-$Au$
(and $d$-$Au$) collisions at center-of-mass energies between 63 and 
200~AGeV~\cite{phenix04,brahms04,phobos04,star05}. 
The emerging consensus is that thermalized 
bulk matter at energy densities ($\sim$20GeV/fm$^3$) well above the 
critical one ($\sim$1GeV/fm$^3$)~\cite{KL03} is being produced
in central $Au$-$Au$ collisions at $\sqrt{s}$=200AGeV, see, \eg,
Ref.~\cite{Ra04qm} and references therein.
However, the microscopic properties of the produced medium are 
expected to be thoroughly probed only once 
electromagnetic (e.m.) and heavy-quark observables become available with
high accuracy. In this paper, we constrain ourselves to the former.

The valuable information contained in e.m. spectra from 
heavy-ion reactions has clearly been demonstrated at the CERN 
Super-Proton-Synchrotron (SPS):  
dilepton and photon spectra from (semi-) central 
$Pb$(158AGeV)-$Pb$/$Au$ collisions have revealed remarkable excess 
radiation over baseline sources (such as initial hard $N$-$N$ 
collisions and final-state meson decays)~\cite{ceres98,na50-00,wa98-00}. 
Theoretical analyses lead to the following conclusions (see, \eg, 
Refs.~\cite{RW00,Alam01,GH03,BR04,Ra04ph} for recent reviews): 
(i) low-mass ($M$$<$1GeV) dilepton spectra require substantial 
medium modifications of the $\rho$-meson in hot and dense hadronic 
matter, with no definite discrimination between a strong broadening as 
found in hadronic many-body calculations and scenarios based on a 
dropping $\rho$-mass; (ii) the predicted prevalence of baryon-driven 
medium effects has been confirmed experimentally by an increased 
enhancement at lower SPS energies ($\sqrt{s}$=8.7AGeV)~\cite{ceres03};
(iii) the sensitivity to QGP emission is small at low mass (10-15\%);   
(iv) the same thermal source can account for the excess observed
at intermediate dilepton masses (1GeV$<$$M$$<$3~GeV) and in direct 
photon spectra, pointing 
at initial temperatures of 200-250MeV. Thus, overall, the calculations 
compared favorably to available data by the year 2002. 
New data from both WA98~\cite{wa98-03} and 
CERES/NA45~\cite{ceres04} with increased sensitivity to low
transverse momenta and high centrality, respectively, are exhibiting 
excess radiation that could be posing a challenge to theory.
An important question to be addressed here is 
whether these observations are related to an incomplete understanding
of the space-time evolution of central $A$-$A$ collisions, or if new 
mechanisms in the microscopic production mechanisms (emission rates)
are required.    

The remainder of this paper is organized as follows. In 
Sec.~\ref{sec_rates} I will give a very brief (and incomplete) survey 
of theoretical frameworks for evaluating e.m. emission rates 
(see Ref.~\cite{Ra04taos} for a more detailed recent account), 
and recall possible connections to chiral symmetry restoration.
Sec.~\ref{sec_spec} contains an analysis of the recent 
measurements of e.m. spectra at the SPS, and an outlook for RHIC. 
Conclusions can be found in Sec.~\ref{sec_concl}.

\section{E.M. Emission Rates and Vector Mesons in Medium}
\label{sec_rates}  
To leading order in $\alpha_{em}=1/137$, production rates of 
photons ($M$=0) and dileptons ($M$$>$0) from a thermal
medium are directly related to the (imaginary part of the) 
e.m. current correlation function (or photon selfenergy) via
\beq
q_0\frac{dR_\gamma}{d^3q} = 
-\frac{\alpha_{\rm em}}{\pi^2}~f^B~{\rm Im}\Pi_{\rm em}(M\!\!=\!\!0)
 \ , \quad  
\frac{dR_{e^+e^-}}{d^4q}= -\frac{\alpha_{\rm em}^2}{M^2\pi^3}~
       f^B~{\rm Im}\Pi_{\rm em}(M\!\!>\!\!0)
\label{rate}
\eeq
($f^B$: Bose distribution), respectively.
In the vacuum, Im$\Pi_{em}$($M$=0)=0, whereas Im$\Pi_{em}$($M$$>$0) can 
be determined by $e^+e^-\to hadrons$, being characterized by roughly 
2 regimes: for  $M$$>$$M_{dual}$$\simeq$1.5GeV, the perturbative 
$q\bar q$ continuum reproduces the data within $\sim$20\%, 
while for $M$$<$$M_{dual}$, Im$\Pi_{em}$ is saturated 
by the vector mesons $\rho^0$, $\omega$ and $\phi$.  Since 
$\Gamma_{\rho\to ee}$$\simeq$
11$\Gamma_{\omega\to ee}$$\simeq$
5.5$\Gamma_{\phi\to ee}$~\cite{pdg04}, 
the isospin-1 ($\rho$) channel
largely dominates the thermal rate. This provides favorable
conditions in the quest for chiral
symmetry restoration, as discussed below. 

In the QCD vacuum, the (approximate) $SU(2)_L\otimes SU(2)_R$ 
chiral symmetry of the QCD Lagrangian is spontaneously 
broken by the formation of a (scalar) quark condensate, 
$\langle \bar q q\rangle$$\simeq$(-250MeV)$^3$. While the 
condensate is not an observable, the spontaneous breaking of chiral
symmetry (SBCS) manifests itself in the excitation spectrum of
the vacuum through the (mass-) splitting of ``chiral partners".  
Most prominent examples in the mesonic sector are the 
non-degeneracy of the pseudo-/scalar ($\pi$-``$\sigma$") 
and axial-/vector ($\rho$-$a_1$) channels (in the
2-flavor case, the $\omega$ is a chiral singlet). Above the
critical temperature for chiral restoration, $T_c$, 
the spectral densities within a chiral multiplet become
degenerate. Thus, medium modifications of the $\rho$-meson
spectral function are hoped to illuminate mechanisms for
chiral restoration. It is clear, however, that this requires  
to establish connections to the $a_1$-channel, as encoded, \eg,
in the second Weinberg sum rule~\cite{Wei67,KS94},
\beq
f_\pi^2 = - \int \frac{ds}{\pi s} ({\rm Im}\Pi_V - {\rm Im}\Pi_A) \ ,
\eeq   
with axial-/vector correlators 
Im$\Pi_{V,A}$=$(m_{\rho,a_1}^4/g_{\rho,a_1}^2)$Im$D_{\rho,a_1}$
using vector dominance (cf.~also Ref.~\cite{UBW02}). One should note 
that other realizations of SBCS are possible; \eg, in 
Ref.~\cite{HY03}, employing Hidden Local Symmetry (HLS) to 
introduce vector mesons into a chiral Lagrangian, the chiral partner 
of the pion has been identified with the longitudinal component of the 
$\rho$-meson field, the so-called ``vector manifestation"
of chiral symmetry. The bare parameters of the effective Lagrangian
($m_\rho^{(0)}$, $F_\pi$, $g_\rho$) are constrained by matching the
1-loop expanded axial-/vector correlators to an operator product
expansion (OPE, for spacelike momenta) at a typical scale of 
$\Lambda$$\simeq$1.1GeV, with no explicit $a_1$ and ``$\sigma$" degrees 
of freedom. The bare couplings and masses are then evolved to the
on-shell points rendering a satisfactory phenomenology of free decays. 

Intense efforts have been devoted in recent years to evaluate
modifications of the vector-meson spectral densities in hot and dense
hadronic matter, employing mean-field models, finite-temperature chiral 
loop expansions, or hadronic many-body theory (see Ref.~\cite{RW00} 
for a review).
The reliability of the calculations resides to a 
large extent on the ability to constrain effective 
interaction vertices by both symmetry principles (gauge and chiral) 
and experimental vacuum decay branchings or scattering data   
(\eg, $\pi N\to\rho N$ or $\gamma N$, $\gamma A$ absorption spectra).  
Information on vector-meson interactions with excited resonances (\eg, 
$\rho$-$\Delta(1232)$) is rather limited. However, they are potentially 
important once thermal abundances are 
significant (\eg, at $T$=170~MeV, $N$ and $\Delta$ densities are 
approximately equal).     

\begin{figure}[thb]
\vspace*{-2.3cm}
\leavevmode
\epsfysize=6.8cm
\epsfbox{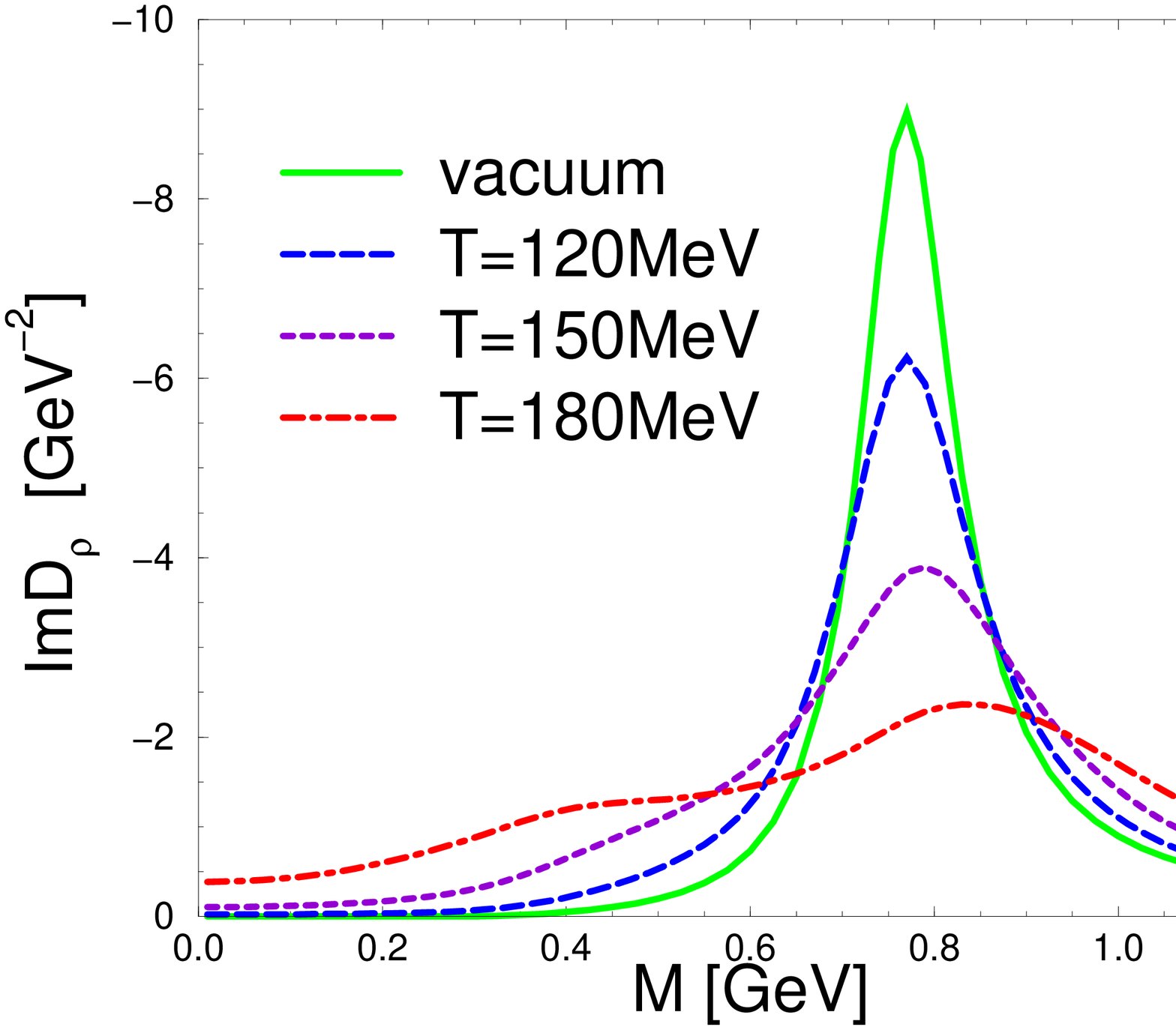}
\hspace*{0.5cm}
\epsfysize=7.4cm
\epsfbox{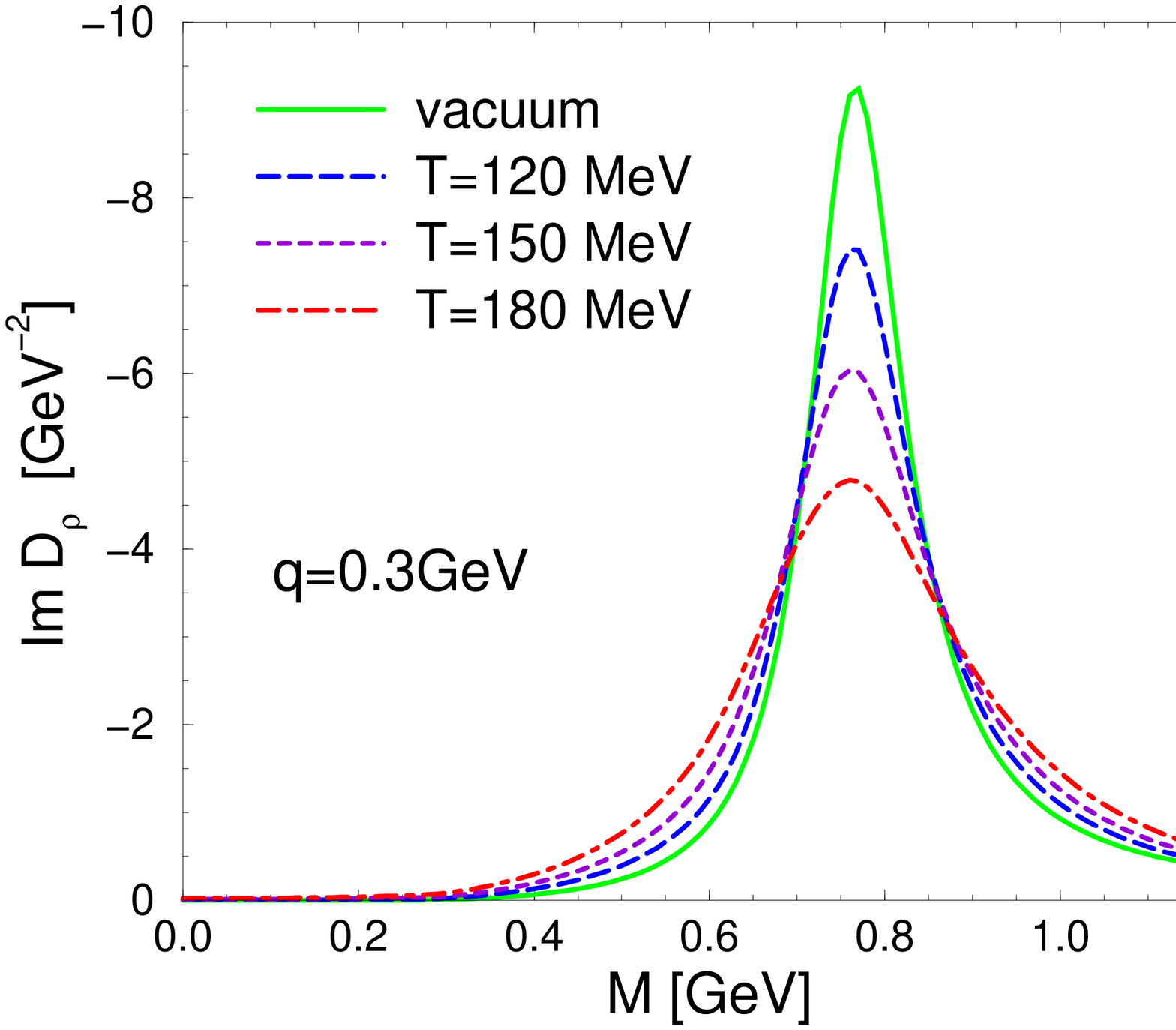}
\vspace*{-0.8cm}
\caption[]{$\rho$-meson spectral functions within hadronic
many-body theory~\protect\cite{RW99}; left panel: hot+dense matter
at temperature [MeV] and baryon density [$\rho_0$]
($\rho_0$=0.16fm$^{-3}$) of [$T$,$\rho_B$]=
[0,0] (solid line), [120,0.1] (long-dashed line), [150,0.7] (short-dashed
line) and [180,2.6] (dash-dotted line),
right panel: hot meson gas.}
\vspace{-0.1cm}
\label{fig_arho}
\end{figure}
In the vector-manifestation scenario~\cite{HY03}, it has been argued 
that two types of medium effects should be accounted for~\cite{HS02}: 
(i) the standard thermal loop corrections affect the $\rho$-mass
only to order ${\cal O}(T^4)$, consistent with earlier findings;
(ii) from matching the thermal correlators to the in-medium OPE    
(with $T$-dependent quark condensates), it has been inferred
that in addition to (i) a reduction of the bare parameters (masses and 
couplings) is required. This, in particular, leads to
a dropping of the $\rho$ (pole-) mass, consistent with the Brown-Rho
scenario~\cite{BR04}, and a vanishing of the vector-dominance coupling
at $T_c$.
 
\begin{figure}[thb]
\vspace*{-0.3cm}
\begin{center}
\hspace*{-1.5cm}
\epsfysize=8.1cm
\epsfbox{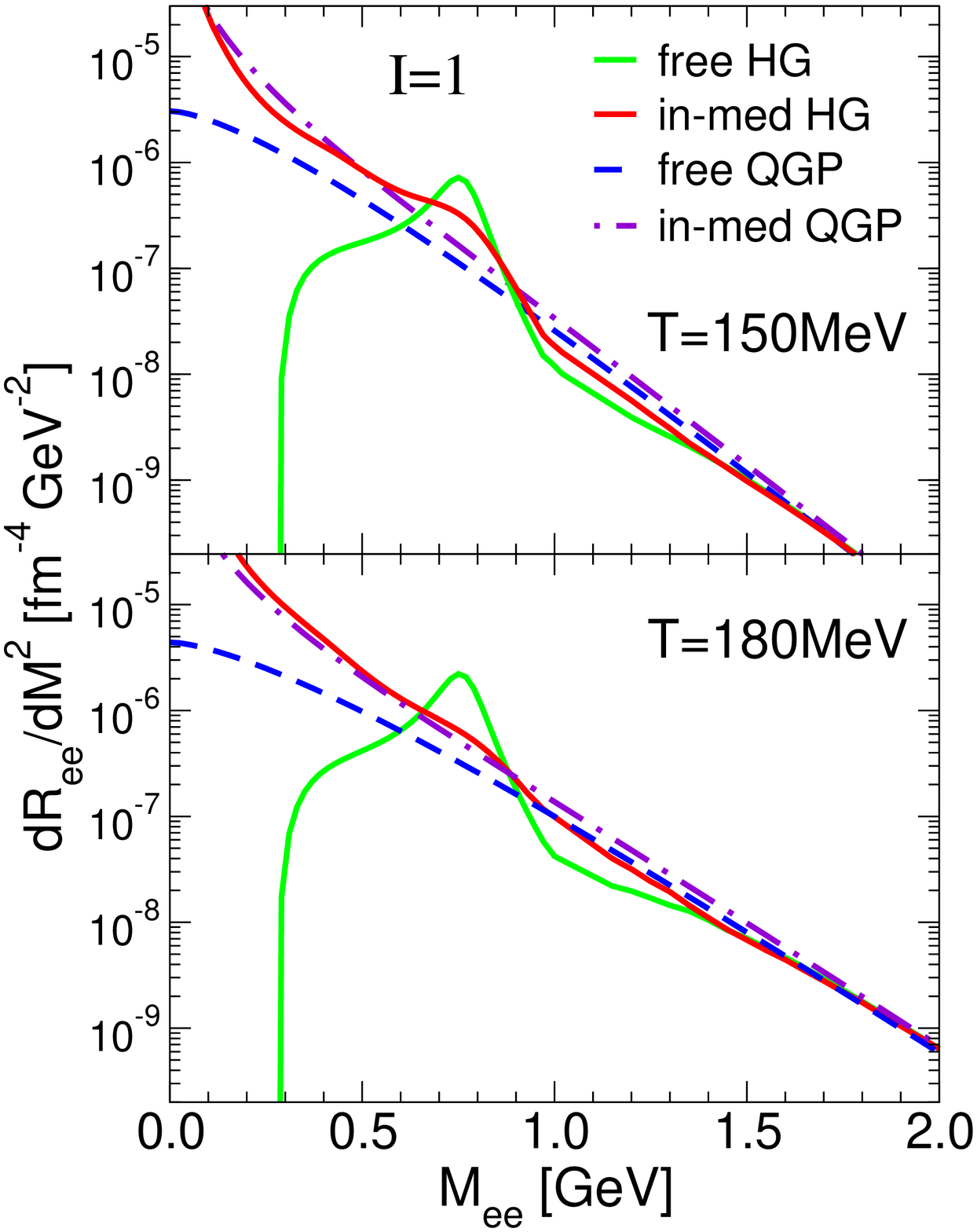}
\hspace{-0.7cm}
\epsfysize=6.1cm
\epsfbox{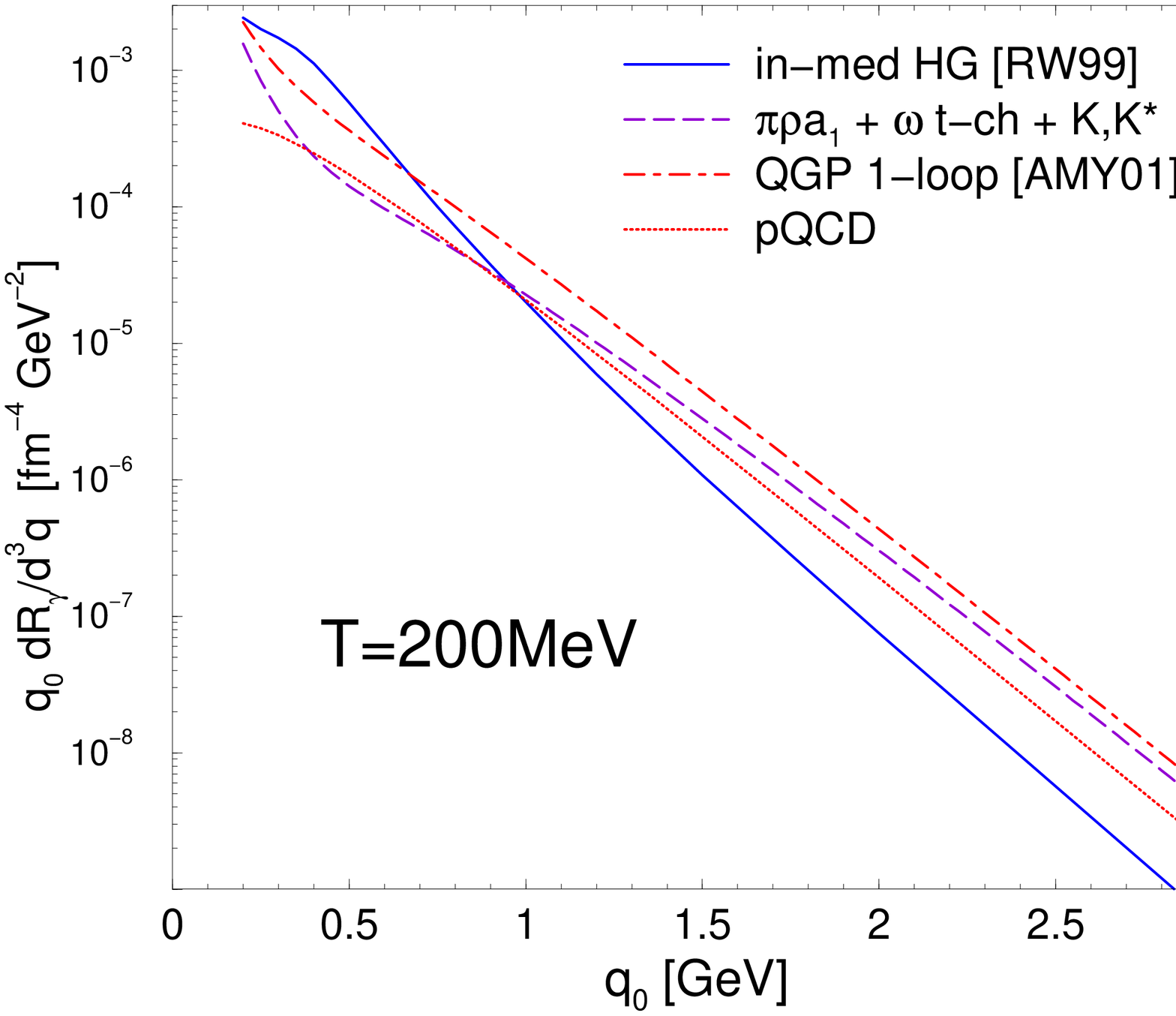}
\end{center}
\vspace*{-1.0cm}
\caption[]{E.m.~emission rates from strongly interacting matter; left 
panel: dileptons from the isovector channel (dashed and dot-dashed 
line: free and hard-thermal-loop (HTL)~\protect\cite{BPY90} improved 
$q\bar q$ annihilation (QGP), solid lines: free and in-medium hadron 
gas); right panel: photons (dotted line: tree-level 
pQCD~\protect\cite{KM81}, dash-dotted: complete leading order 
QGP~\protect\cite{AMY01}, solid: hadronic many-body contributions at 
the photon point; dashed: leading order meson-exchange 
reactions~\protect\cite{TRG04}).}
\label{fig_rates}
\vspace{-0.1cm}
\end{figure}

A typical example of a hadronic many-body calculation of the $\rho$-meson
propagator, $D_\rho=[M^2-(m_\rho^{(0)})^2-\Sigma_\rho]^{-1}$, is shown in 
Fig.~\ref{fig_arho} (including conservative
estimates of contributions from higher resonances)~\cite{RW99}.  
The essential features, also shared by other calculations of similar
kind~\cite{Post01,Cabrera02,Lutz02}, are: 
(1) strong broadening of the resonance structure due to 
large (negative definite) imaginary parts of the in-medium selfenergy, 
$\Sigma_{\rho}$; (2) little mass shifts due to cancellations in the real
part of $\Sigma_{\rho}$; (3) prevalence of baryonic over mesonic
medium effects (this remains true for net-baryon free matter 
as long as the total baryon density, 
$\varrho_{B,tot}$=$\varrho_B$+$\varrho_{\bar B}$, is significant).
The impact of baryons is especially pronounced in the regime below the 
free $\rho$-mass, which in thermal dilepton rates is augmented  
by Bose- and $M^{-2}$-factors (due to the photon propagator), 
cf.~Eq.~(\ref{rate}). Three-momentum integrated dilepton rates from 
both hadronic gas (HG) and QGP are compiled in the left panel of 
Fig.~\ref{fig_rates}. One finds that both in-medium rates are strongly 
enhanced over their free counterparts at low $M$, a typical   
many-body effect (and necessary to have a nonzero photon rate for 
$M$$\to$0). In fact, extrapolating HG and QGP rates (up/down) to 
$\sim$$T_c$, rather close agreement emerges, \ie, the ``matching" is 
automatic! Given the importance of baryonic effects, it would be
interesting to see whether a more complete treatment of hadronic
interactions within the vector manifestation scenario still 
mandates an ``intrinsic" $T$-dependence of the bare parameters.

Turning to photon production rates (Fig.~\ref{fig_rates}, right panel), 
the complete leading-order in 
$\alpha_s$ result has recently been obtained in Ref.~\cite{AMY01};
a global enhancement of about a factor 3 over early tree-level 
estimates~\cite{KM81} has been found.
In the hadronic sector, consistency between low-mass dileptons
and photon rates has recently been put forward~\cite{TRG04}.  
The processes most relevant for the low-mass enhancement around
$M$$\simeq$0.5GeV dominate the photon rate up to energies of 
$\sim$1GeV. Above, $t$-channel meson exchanges take over, with
formfactor effects reducing the
emission strength at high energies appreciably
(by a factor $\sim$3 at $q_0$$\simeq$3GeV). In the vicinity of 
$T_c$, in-medium QGP and HG rates are again quite comparable 
in strength.

\section{Electromagnetic Spectra in Heavy-Ion Collisions}
\label{sec_spec}
To calculate e.m. spectra in heavy-ion collisions, the emission
rates are to be convoluted over the space-time history of the 
reaction, assuming local
thermal equilibrium. The selection of results discussed below is
based on a simple expanding (isentropic and isotropic) fireball model 
which is consistent with measured particle 
yields and radial flow. The resulting photon spectra are, \eg, quite 
consistent with more elaborate hydrodynamic 
simulations~\cite{HRR02,Sriv05}, see also Ref.~\cite{Ra04ph}.
The equation of state for the fireball includes finite 
meson-chemical potentials ($\mu_\pi$, $\mu_K$, etc.) after chemical 
freezeout, entailing slightly steeper e.m. spectra (due to 
faster cooling but with enhanced yields), which is 
particularly relevant at low $M$ and $q_t$.  

\subsection{Photons}
\label{ssec_phot}
\begin{figure}[tbh]
\begin{center}
\vspace*{-1.7cm}
\hspace{-1.7cm}
\epsfxsize=5.6cm\epsfbox{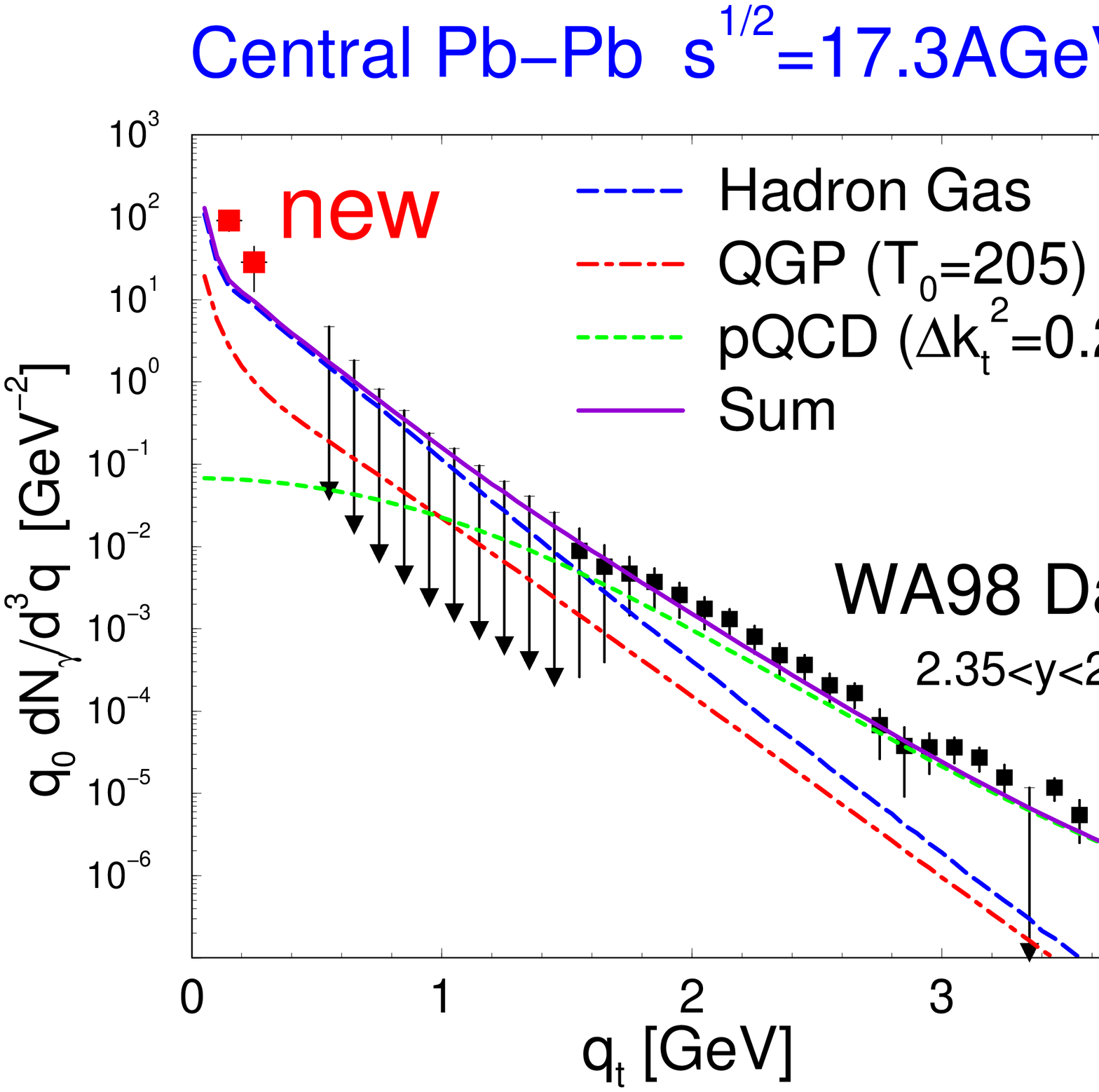}
\hspace*{0.4cm}
\epsfxsize=5.2cm\epsfbox{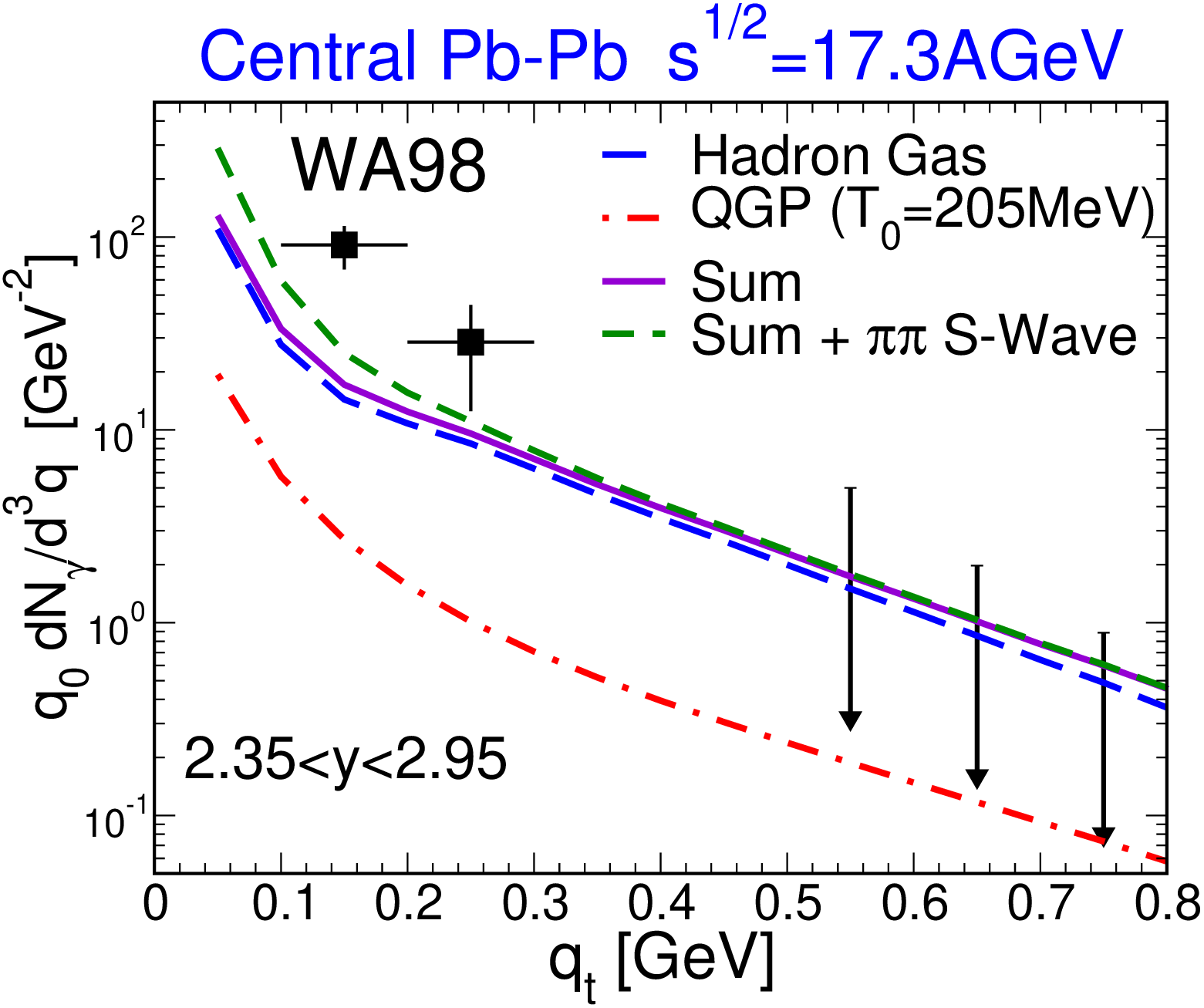}
\end{center}
\vspace*{-0.8cm}
\caption[]{Direct Photon Spectra as measured in central 
$Pb$(158AGeV)-$Pb$ collisions~\protect\cite{wa98-00,wa98-03} compared 
to calculations of 
Ref.~\protect\cite{TRG04} (left panel). The right panel magnifies the 
low-momentum region with the short-dashed line including additional 
contributions from Bremsstrahlung off free $S$-wave $\pi\pi$ 
scattering~\protect\cite{LTRG05}.}
\label{fig_phot}
\vspace{-0.1cm}
\end{figure}
The left panel of Fig.~\ref{fig_phot} shows a comparison of WA98 data
from central $Pb$-$Pb$ at the SPS with a calculation based on the
hadronic~\cite{TRG04} and complete-LO QGP~\cite{AMY01} photon rates 
displayed in Fig.~\ref{fig_rates} (right panel). 
The main points to be made are: 
(i) prevalence of the primordial contribution from $p$-$p$ 
collisions above $q_t$$\simeq$1.5GeV once a moderate Cronin effect 
(estimated from $p$-$A$ data) is accounted for, 
implying that "standard" initial temperatures ($\bar T_0$$\simeq$205MeV
corresponding to a formation time $\tau_0$=1fm/c) for the thermal yield
are in line with the data; (ii) consistency with thermal 
dileptons~\cite{RW00,Ra04taos} both in terms of emission rate and 
space-time evolution; (iii) underestimation of the new low-$q_t$ data
by the theory predictions~\cite{TRG04,Sriv05}. The latter
slightly improve upon inclusion of Bremsstrahlung
off $S$-wave $\pi\pi\to\pi\pi$ scattering. This opens the exciting 
possibility that medium effects in the scalar-isoscalar (``$\sigma$")
channel are at the origin of the enhancement. However, before 
conclusions in that direction can be drawn, a thorough evaluation of 
coherent Bremsstrahlung from the in- and outgoing hadronic charges is 
mandatory~\cite{LTRG05}.

\subsection{Dileptons}
\label{ssec_dilep}
\begin{figure}[htb]
\begin{center}
\vspace*{-1.7cm}
\hspace*{-0.8cm}
\epsfxsize=5.6cm\epsfbox{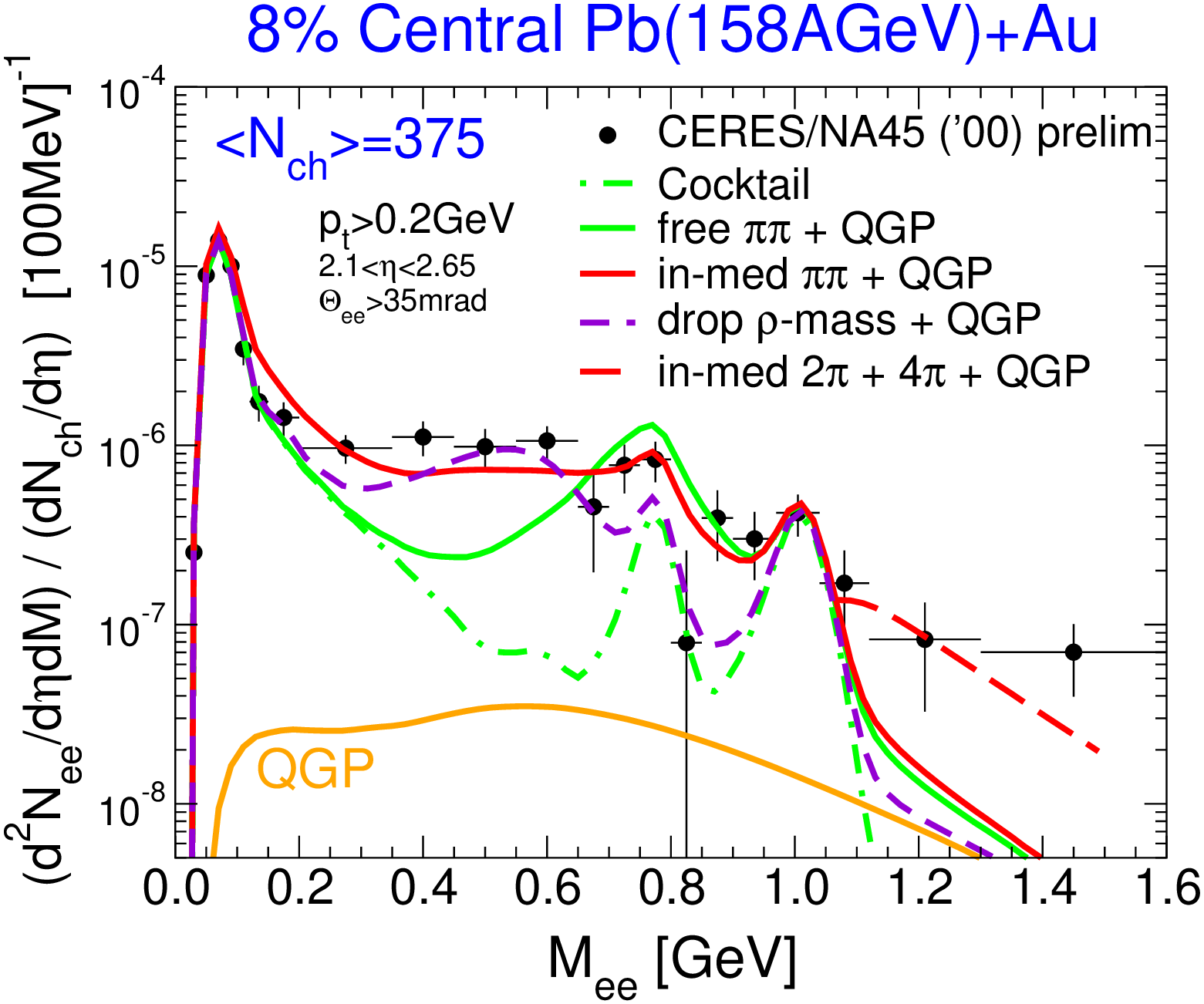}
\hspace*{0.6cm}
\epsfxsize=5.6cm\epsfbox{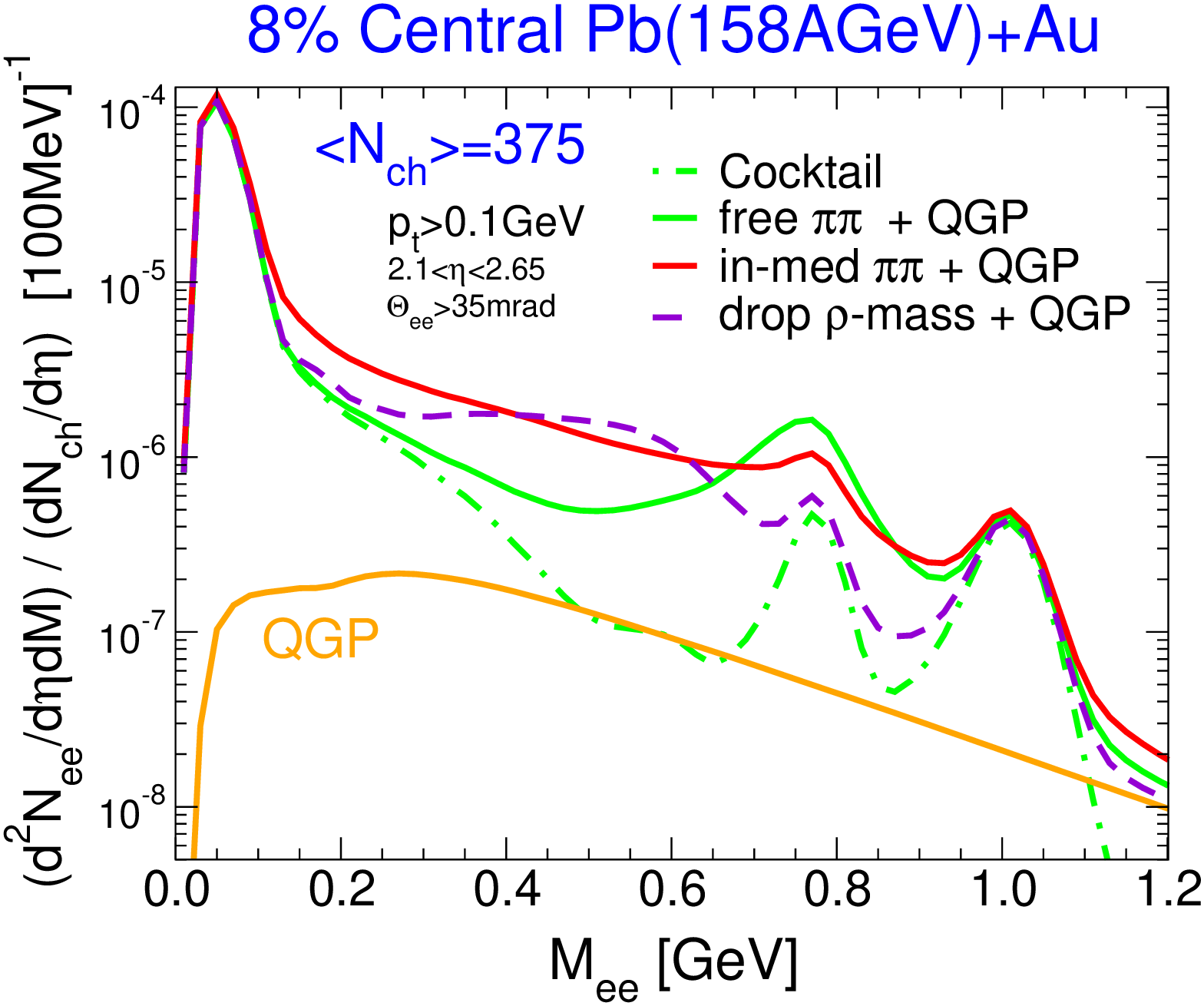}
\end{center}
\vspace*{-1cm}
\caption[]{Left panel: preliminary CERES data (normalized to the $\pi^0$ 
decay) with single-$e^\pm$ cuts $p_t$$>$0.2GeV~\protect\cite{ceres04};
theoretical curves~\protect\cite{RW00}, added to the final-state decay 
cocktail (dash-dotted line)~\protect\cite{ceres04}, are calculated from 
an expanding fireball with a HG yield using a $\rho$ spectral function 
with either vacuum shape (green solid line), a dropping mass 
(short-dashed line), or many-body effects (red solid line), 
plus a HTL-QGP yield~\protect\cite{BPY90}; 
the long-dashed line for $M$$>$1.1GeV accounts for contributions of 
4-$\pi$ states to the e.m.~correlator in the hadronic phase. 
Right panel: theoretical predictions for a reduced single-$e^\pm$ cut 
of $p_t$$>$0.1GeV (line identification as in the left panel), with the 
decay cocktail according to Ref.~\protect\cite{ceres05}.}
\label{fig_ceres}
\vspace{-0.1cm}
\end{figure}

In the year 2000, CERES/NA45 measured low-mass $e^+e^-$ with 
improved statistics and mass resolution over previous 
results for central $Pb$(158AGeV)+$Au$ collisions, 
see left panel of Fig.~\ref{fig_ceres}~\cite{ceres04}. 
The spectra (with the standard single-$e^\pm$ cut $p_t$$>$0.2GeV) are
compared to theory predictions for somewhat lower centrality 
($N_{ch}$=250), upscaled by the charged-particle 
multiplicity (i.e., a factor 375/250; this may slightly overestimate an 
explicit calculation at $N_{ch}$=375). On the one hand, below the free
$\rho$-mass, conclusions from earlier data are confirmed 
(cf.~Introduction). On the other hand, one now has better sensitivity 
to the mass regions between the $\omega$ and the $\phi$, and above 
the $\phi$. For the former, the (broadenend) many-body spectral
function appears to be favored, although it is not clear whether in 
the dropping-mass scenario an almost vanishing strength of the 
e.m.~correlator is realistic. Above the $\phi$-mass, the original 
calculations~\cite{RW99} underestimate 
the data. However, these were based on contributions from 2-pion states
only, whereas the free e.m.~correlator is known to be dominated by  
4-pion (and higher) states above $M_{ee}$$\simeq$1.2GeV~\cite{Aleph98}.
In addition, vector-axialvector mixing~\cite{DEI90,Chan99} is expected
to enhance Im$\Pi_{em}$ in the mass region between 1 and 1.5GeV. To 
schematically include these effects, we have implemented the (``dual") 
HTL-improved pQCD rate for the hadronic emission above the $\phi$-mass 
(cf.~Fig.~\ref{fig_rates}) with a pertinent  4-pion fugacity factor, 
e$^{4\mu_\pi/T}$. We have explicitly checked that contributions from 
Drell-Yan annihilation are small (below the QGP yield), as is also 
expected for open-charm decays~\cite{BMDL98}.
The approximate agreement with the CERES data above 1GeV reconfirms the
consistency with earlier calculations~\cite{RS00} in the context
of the excess observed by NA50~\cite{na50-00}.

In view of the low-$q_t$ photon enhancement observed by WA98 and 
its possible connections to the timelike regime, 
it is of interest to further probe dileptons at low momentum.
In the right panel of Fig.~\ref{fig_ceres} theoretical predictions
are given for a reduced single-$e^\pm$ $p_t$-cut of 0.1~GeV. 
The most relevant mass range is identified as $M$$\simeq$0.2-0.4GeV, 
where thermal radiation is still prevalent (below, $\eta$ and $\pi^0$ 
Dalitz decays take over) and generates a factor $\sim$2-3 enhancement
over the spectra with $p_t$$>$0.2~GeV (at least for the in-medium
many-body spectral function). An analysis of the 1996 CERES data
with $p_t$$>$0.1~GeV for 32\% central collisions~\cite{Hering02}, 
albeit with rather large statistical errors, is in approximate agreement 
with these calculations, but more precise data would be very valuable.

\subsection{RHIC}
\label{ssec_rhic}
\begin{figure}[tb]
\vspace*{-0.3cm}
\hspace{0.5cm}
\leavevmode
\epsfxsize=5.6cm\epsfbox{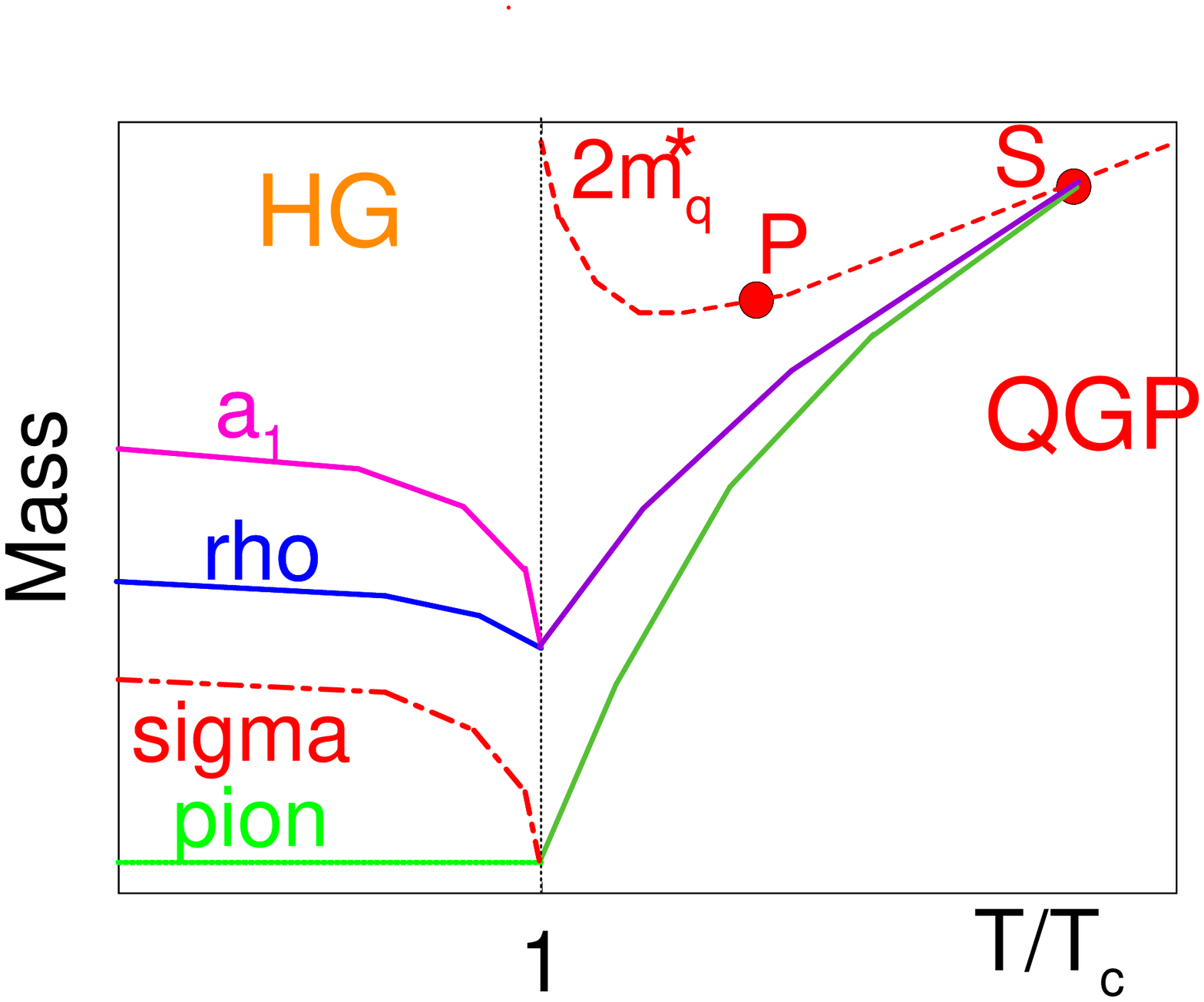}
\hspace*{0.5cm}
\epsfxsize=5.6cm\epsfysize=5cm\epsfbox{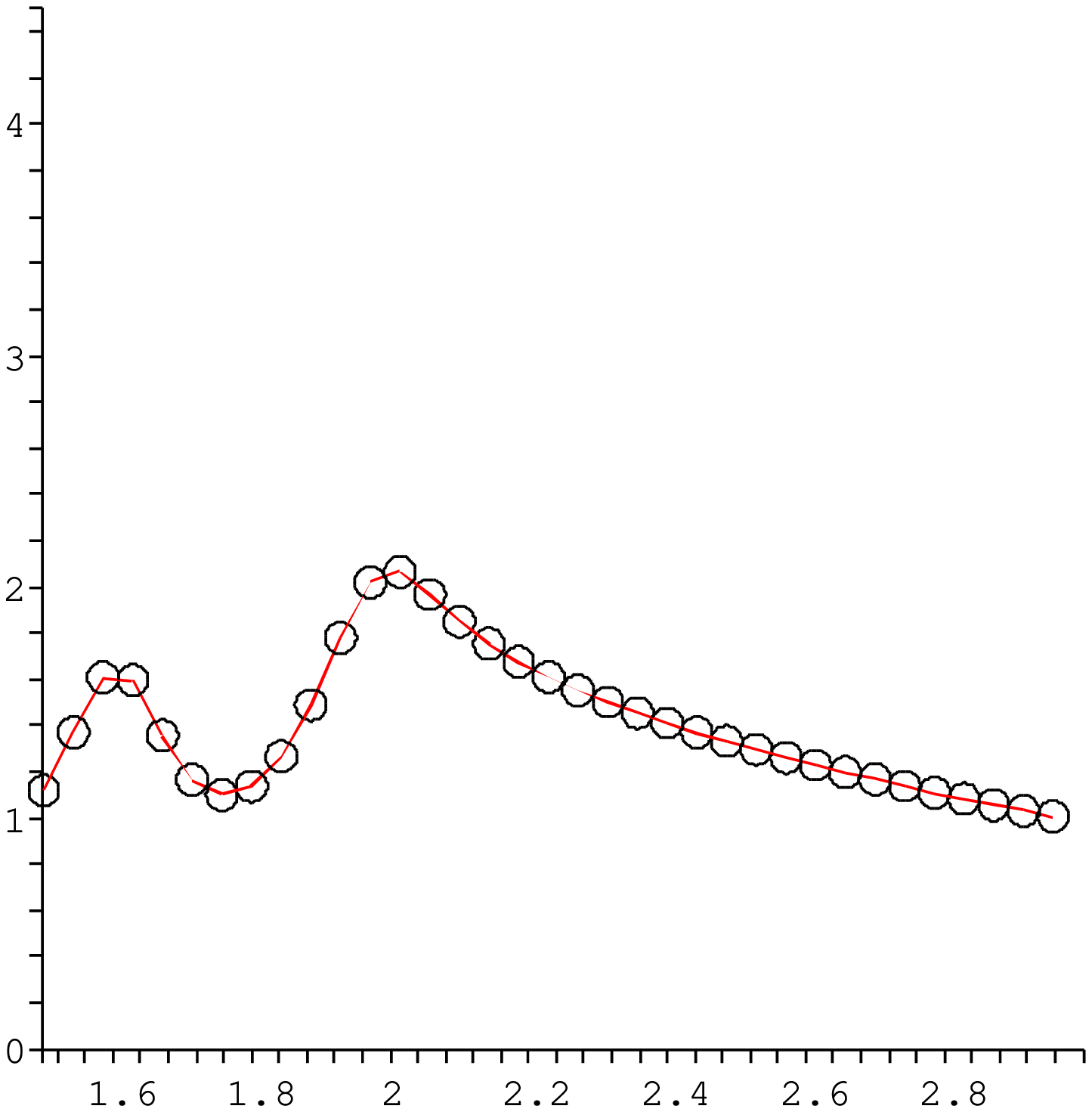}
\vspace*{-0.3cm}
\caption[]{Left panel: schematic dependence of hadronic masses on 
temperature~\protect\cite{SZ03}. Right panel: ratio of the 
time-integrated QGP dilepton yield at RHIC from $\rho$-like bound 
states~\protect\cite{CS05} to ``standard" $q\bar q$ 
annihilation~\protect\cite{Ra02}. The two bumps
arise from approximately constant-mass $\rho$-mesons along the
"zero-binding line" (higher mass) and from the mixed phase
(lower mass). The $x$-axis denotes invariant dilepton mass in
units of the quark quasiparticle mass in the QGP.}   
\label{fig_res}
\vspace{-0.1cm}
\end{figure}

Let us finally turn to prospects for RHIC. Due to copious production
of open-charm pairs, their correlated semileptonic decay will become
a major component in the dilepton spectrum. Nevertheless, due
to  medium effects on the $\rho$ meson comparable to those at the SPS, 
the thermal radiation at low mass is expected to outshine
the open-charm contribution by a factor $\sim$2-3~\cite{Ra02}. 
Starting from just above the $\phi$-mass, correlated charm decays
as extrapolated from $p$-$p$ spectra~\cite{Aver01} dominate over
thermal radiation. This situation could change, if
(a) charm quarks thermalize in the QGP leading to a softening 
in the pertinent dilepton $M$-spectra, and/or (b) the thermal
emission rate becomes enhanced. A possible mechanism for
the latter has been analyzed in Ref.~\cite{CS05} in terms of 
``$\rho$"-resonances surviving in the QGP~\cite{SZ03} (cf. left panel
of Fig.~\ref{fig_res}), for $T$$\le$2$T_c$ or
so. Depending on their width, an excess of up to a factor 2 over the
perturbative QGP emission occurs in the integrated spectra, see
right panel of Fig.~\ref{fig_res}.    

\section{Conclusions}
\label{sec_concl}
Electromagnetic probes are providing exciting insights into
the properties of hot and dense matter, as produced in relativistic
heavy-ion collisions. 
The theoretical objective is to achieve an accurate enough calculation 
of the e.m. correlation function that allows a consistent explanation
of thermal photon and dilepton spectra. This has to be coupled with 
a description of the interacting many-body system as a whole, to enable 
the identification of the relevant degrees of freedom and their
chiral properties. In the latter context, the measurement of
$\pi^\pm\gamma$ invariant-mass spectra, with the idea to extract
spectral properties of the $a_1$, would be highly valuable.
 
Current hadronic calculations exhibit medium effects strong enough
to essentially ``melt" the light vector mesons ($\rho$, $\omega$)
toward the (pseudo-) phase boundary, rendering a ``matching" of the 
e.m. correlator to structureless pQCD results for the QGP. It will be 
important to find out in how far nonperturbative interactions
in the QGP modify this picture, and at which mass scale they appear. 

At the SPS, new dilepton data are so far in reasonable agreement 
with medium-modified $\rho$ spectral functions, whereas low-momentum 
photon data indicate an appreciable excess beyond current theoretical 
expectations. Rather general arguments
based on the interplay of temperature and space-time volume relegate 
the origin of the latter effect to the (late) hadronic phase of
central $A$-$A$ collisions. A softened $\sigma$-mode might be
a promising candidate to explain these observations. 
Work in this direction is in progress.

\section*{Acknowledgments}
I thank U.~Mosel for the invitation to a very exciting 
European Graduate School workshop in Gie{\ss}en, 
and J. Wambach for the hospitality
during my stay at TU Darmstadt. This work was supported in part
by a U.S. National Science Foundation CAREER award under grant
PHY-0449489. 
 

\vfill\eject
\end{document}